\documentclass[superscriptaddress,aps,pra,10pt]{revtex4-2}

\usepackage{amsmath, amssymb}
\usepackage{siunitx}
\usepackage{lineno}
\usepackage{graphicx}
\usepackage[hidelinks]{hyperref}
\begin{document}

\title{Correcting for objective sample refractive index mismatch in extended field of view selective plane illumination microscopy}

\author{Steven J. Sheppard}
\author{Peter T. Brown}
\author{Douglas P. Shepherd}
\email{douglas.shepherd@asu.edu} 
\address{Center for Biological Physics and Department of Physics, Arizona State University, Tempe, AZ}

% use {asbstract*} to suppress the copyright line. Copyright information will be added in production
\begin{abstract}
Selective plane illumination microscopy (SPIM) is an optical sectioning imaging approach based on orthogonal light pathways for excitation and detection. The excitation pathway has an inverse relation between the optical sectioning strength and the effective field of view (FOV). Multiple approaches exist to extend the effective FOV, and here we focus on remote focusing to axially scan the light sheet, synchronized with a CMOS camera's rolling shutter. A typical axially scanned SPIM configuration for imaging large samples utilizes a tunable optic for remote focusing, paired with air objectives focused into higher refractive index media. To quantitatively explore the effect of remote focus choices and sample space refractive index mismatch on light sheet intensity distributions, we developed a computational model integrating ray tracing and field propagation. We validate our model's performance against experimental light sheet profiles for various SPIM configurations. Our findings indicate that optimizing the position of the sample chamber relative to the excitation optics can enhance image quality by balancing aberrations induced by refractive index mismatch. We validate this prediction using a homebuilt, large sample axially scanned SPIM configuration and calibration samples.
\end{abstract}

\maketitle

%%%%%%%%%%%%%%%%%%%%%%%%%%  body  %%%%%%%%%%%%%%%%%%%%%%%%%%
\section{Introduction}
Selective plane illumination microscopy (SPIM), or light sheet fluorescence microscopy (LSFM), encompasses techniques that enable rapid volumetric imaging of samples. SPIM approaches commonly involve two orthogonal objectives, one delivering light to the sample (excitation pathway) and one collecting emitted fluorescence (emission pathway)~\cite{voie1993,huisken2004,power2017}. Shaping the excitation laser into a ``light sheet" that is thin along the detection axis enables optical sectioning, leading to high-contrast images, even in densely labeled samples. Successful applications of SPIM span a wide range, including imaging of entire zebrafish embryos~\cite{huisken2004}, tracking single cell dynamics in early mouse development\cite{mcdole2018}, observing neural activity in the brains of small organisms~\cite{ahrens2013}, imaging of the central nervous system of mice~\cite{kathe2022}, and extracting the hydrodynamic properties of flagella by observing their free diffusion~\cite{franky2023}. These applications demonstrate SPIM's capability to provide detailed, high-resolution images while minimizing photodamage and photobleaching.
 
All light sheet generation strategies have an intrinsic coupling of the width at the focus and the axial extent of the focus. One strategy to extend the axial extent for a given focus width is to use self-healing beams, such as Bessel, Lattice, or Airy beams~\cite{remacha2020, chang2020, liu2023}. Gaussian beam propagation defines the the beam waist $\omega_0 = \frac{2 \lambda}{\pi NA_{exc}}$ and confocal parameter (twice the Rayleigh length) $2 \times z_R = 2 \times \frac{2 \lambda}{\pi NA^2_{exc}}$ to parameterize the focus width and propagation length. Recently, both simulations and experiments have settled that Gaussian and ``sinc" (given by a flat top profile) beams have larger $z_R$ than the Gaussian prediction, allowing for high-contrast imaging volumes similar to certain self-healing beams~\cite{remacha2020,chang2020}. The theoretical discrepancy arises because Gaussian predictions rely on the paraxial wave equation, which assumes small angles and does not account for diffraction effects at larger angles. Predominantly, prior simulations characterized various static and scanned light sheet modalities by propagating analytic electric field amplitude and phase distributions using beam propagation methods, to generate the 3D intensity distribution about the focal plane~\cite{remacha2020, chang2020, shi2022, deng2021}. Experimentally, deviation from a Gaussian focus results from overfilling the pupil, leading to a more uniform line intensity in the pupil and a $sinc$ rather than a Gaussian distribution at the focus~\cite{liu2023}.

\begin{figure}[ht!]
    \centering
    \includegraphics[width=1.0\linewidth]{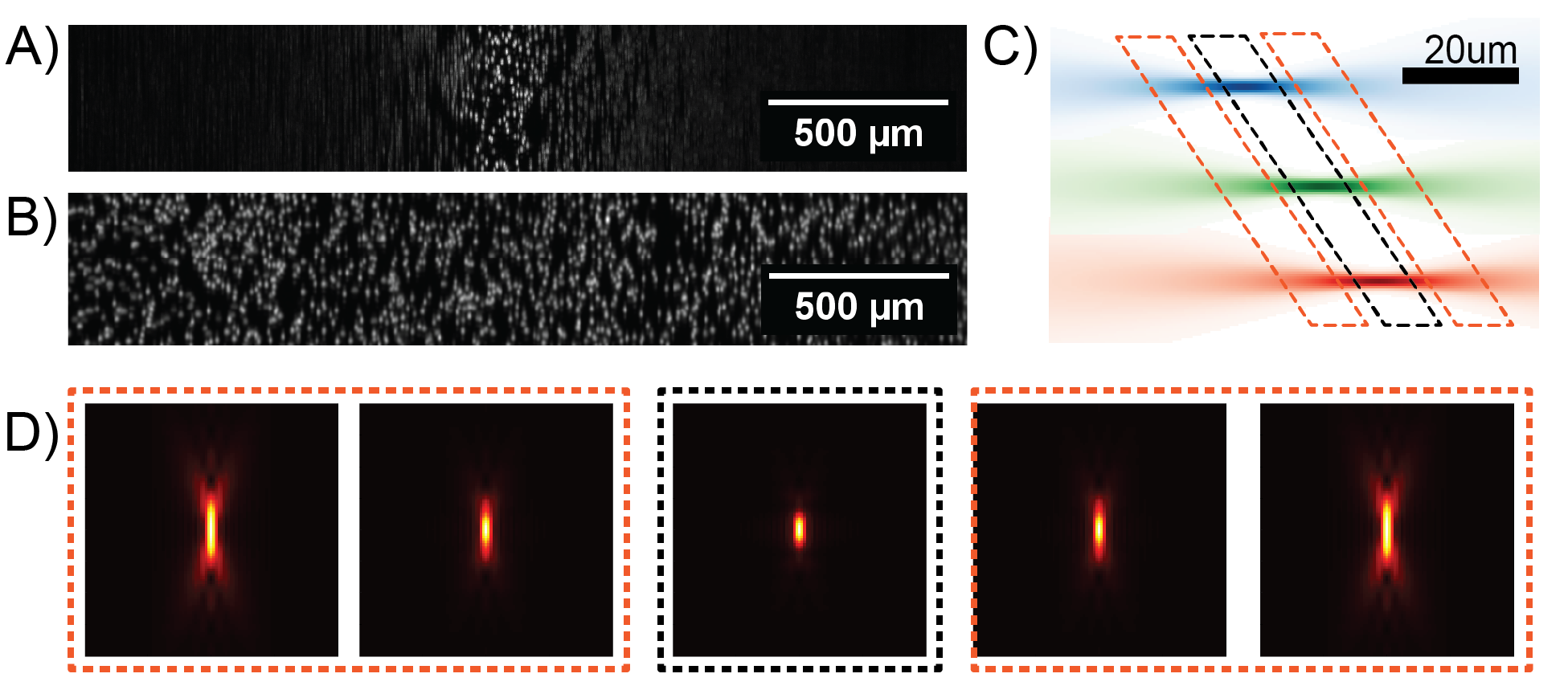}
    \caption{(A-B) \qty{200}{\micro\meter} z-stack (dz=\qty{0.5}{\micro\meter}) ZX maximum intensity projections (gamma=0.75) of  \qty{1}{\micro\meter} beads embedded in agar. (A) Acquired using a static light sheet, illustrating a limited FOV. (B) Acquired using axially-scanned light sheet microscopy to extend the FOV. (C) Illustrating the translation a focus and selection of the in-focus and the out-of-focus regions. (D) Spatially variant PSF originating from various regions of a static light sheet.}
    \label{fig lightsheet}
\end{figure}

For samples that extend beyond the $\omega_0, z_R$ for a given optical configuration, there will be a spatially variant resolution and contrast, Figure~\ref{fig lightsheet}. Two common strategies to achieve a spatially invariant imaging system are to (1) limit the camera field of view (FOV) to the invariant region of the excitation or (2) increase the confocal parameter by increasing the excitation beam's $\omega_0$. The first strategy limits the effective imaging rate, and the second limits the optical sectioning, contrast, and potential resolution gains. 

To fully exploit SPIM's potential space-time bandwidth, several innovative strategies have emerged to create a spatially invariant imaging system by extending the region of ``best" excitation. Conceptually, these approaches synthesize an extended, high-quality FOV by imaging different parts of the sample with a different light sheet, depicted in Figure~\ref{fig lightsheet}. The differences between methods lie in the mode of extension, ranging from physically translating the sample through the light sheet focus~\cite{schacht2010}, acquiring independent images of the light sheet in different positions~\cite{gao2015}, or continuously sweeping the light sheet position via remote focus~\cite{botcherby2008, chmielewski2015}, synchronously with the digital readout of a programmable CMOS detector~\cite{dean2015}.

Inspired by recent insights into light sheet excitation profiles obtained by modeling and experiment, we sought to undertake a more comprehensive effort to model light sheet profiles, subject to remote focusing, in refractive index (RI) mismatched samples. Specifically, we aim to better understand the interplay between the specific remote focusing implementation and heterogeneous RI pathway on the excitation beam's 3D intensity distribution. Heterogeneous RI can arise from mismatches in the excitation objective versus sample chamber, imaging media, sample, or some combination of all of these. To further explore the consequences of instrument or experimental design choices on the resulting image quality, we present here a computational approach to simulate the 3D light sheet intensity distribution that integrates the remote focus pathway aberrations as well as the effects of a converging beam propagating through an index mismatch.

We focus on air immersion excitation objectives paired with a glass chamber and aqueous imaging media, a combination commonly found when imaging large samples. Both our computational and experimental results confirm that low numerical apertures (NA) remote focusing with a tunable optic is a robust method to extend the FOV with uniform image contrast, even in the presence of a RI mismatch. However, even at moderate NA ($\sim 0.3$), we find significant changes to excitation beam focus location and beam intensity profile for tunable optic remote focusing. In the presence of heterogeneous RI, our model predicts that there exists a combination of applied light sheet displacement and excitation lens to sample chamber spacing that minimizes aberrations in the excitation path. We experimentally validated the model predictions by characterizing the PSF across the FOV for static light sheets, demonstrating that there exist optimal combinations of light sheet position and objective-chamber spacing.

\section{Methods}
\subsection{Optical modeling}
To accurately model both aberrations caused by the light sheet optical train and diffraction during beam propagation, we developed a model which incorporates both ray tracing and field propagation. First, we launched a collection of collimated rays from a starting plane. We model the intensity distribution by stochastically sampling rays from either a Gaussian or uniform distribution to simulate Gaussian and flat-top laser profiles. Then, we ray traced through the optical system by applying Snell's law of refraction at each surface while tracking the optical path length (OPL) of each ray. This allows us to account for aberrations induced by the optical system.

We modelled optical elements as flat or spherical surfaces with material refractive indices~\cite{refractiveindex} according to the manufacturers specifications and a propagation wavelength of \qty{561}{\nano \meter}. Objectives are modeled as perfect lenses which transform a ray's height and angle in the front focal plane (FFP) to the angle and height in the back focal plane (BFP) according to
\begin{eqnarray}
\frac{h_2}{f} &=& n_1 \sin \theta_1 \label{eq:perfect_lens_h}\\
n_2 \sin \theta_2 &=& -\frac{h_1}{f} \label{eq:perfect_lens_theta}
\end{eqnarray}
where $f$ is the vacuum focal length and we suppose the lens is immersed in media with refractive index $n_1$ and $n_2$ the left and right respectively. The FFP is $n_1 f$ to the left of the lens, and the BFP is $n_2 f$ to the right of the lens. This construction ensures that an imaging system composed from two perfect lenses satisfies the Abbe sine condition, but does not account for either field-dependent or field-independent aberrations.
% don't add empty line between equation and next sentence, or latex will treat it as a new paragraph and indent

After ray tracing through the final optical surface, we obtain the electric field  on a grid and propagate it through the light sheet focus using the exact-transfer function (angular spectrum) method \cite{goodman2005} that relates the electric field at axial position $z$ to the initial field,
\begin{equation}
    E(x,y,z) = \mathcal{F}^{-1}\left[\mathcal{F}\left[E(x, y, z=0)\right] e^{\imath z \sqrt{\left(\frac{2\pi n }{\lambda}\right)^2 - k_x^2 - k_y^2}}\right].
    \label{eq transferfunction}
\end{equation}
Here, $\mathcal{F}$ represents a 2D Fourier transform over the $x$-, and $y$-coordinates. This approach entails utilizing a scalar approximation which is valid due to the moderately low numerical apertures commonly used in light sheet illumination \cite{olarte2018}.

\begin{figure}[ht!]
    \centering
    \includegraphics[width=1.0\textwidth]{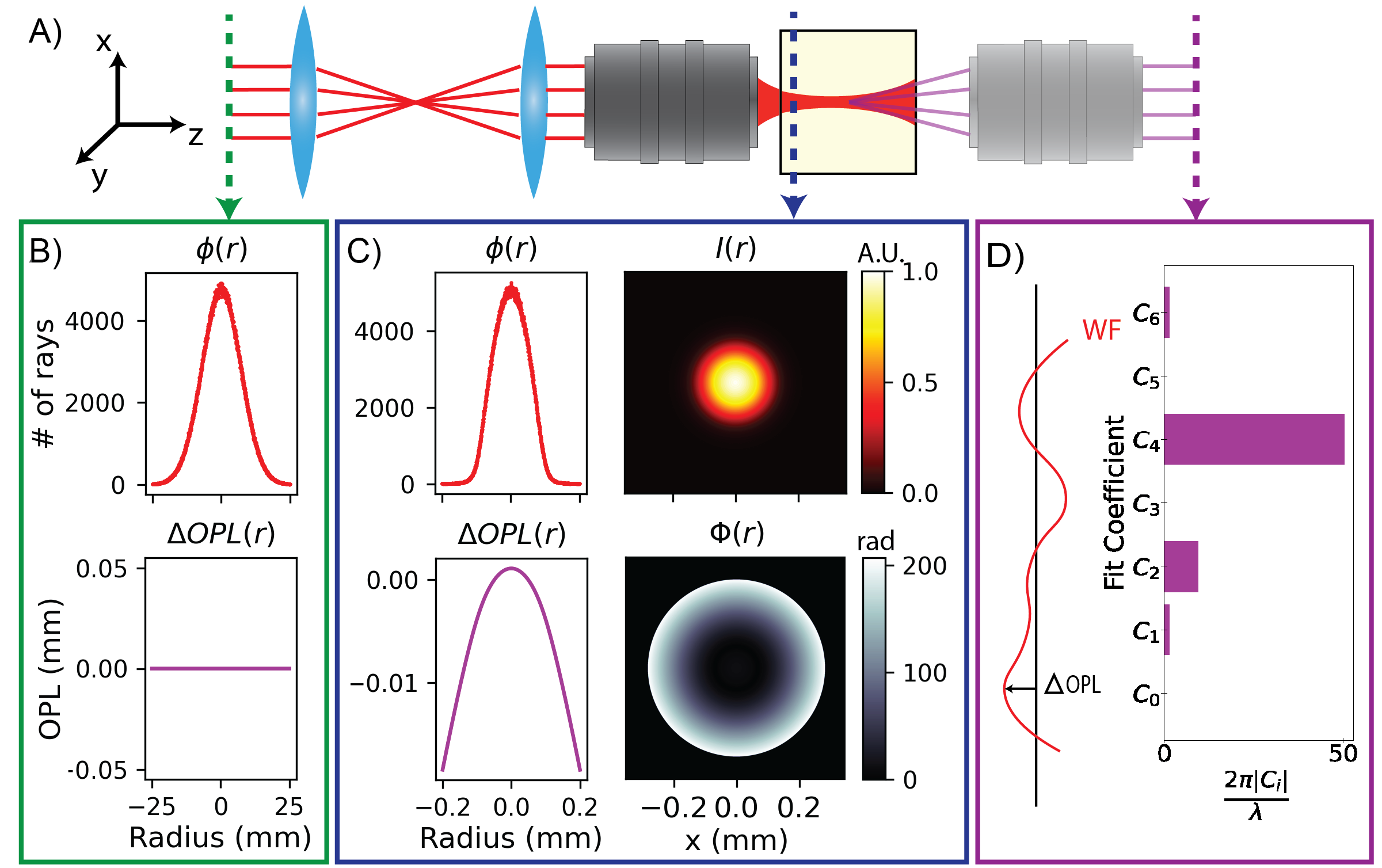}
    \caption{(A) Light sheet model excitation pathway, where \emph{z} is aligned with the propagation direction and \emph{x} along the narrow axis of the light sheet. (B) The initial ray flux, $\phi(r)$, and path length difference, $\Delta OPL(r)$. (C) The ray $\phi(r)$ and $\Delta OPL(r)$ input for computing the electric field to be propagated using the exact-transfer function. The resulting intensity, $I(r)$, and electric field phase amplitude, $\Phi(r)$ are shown. (D) Focus quality was estimated by fitting the wavefront in a pseudo objective's back focal plane.}
    \label{fig model overview}
\end{figure}

For the optical pathways considered here, we assumed radial symmetry and only ray traced a one-dimensional slice through each optical pathway. This gives us a collection of points $\{(r_i, OPL_i)\}_{i=1, ..., N}$ (see fig.~\ref{fig model overview}C). From these discrete points, we create an interpolating function for the optical path length, $OPL_\text{int}(r)$. 
To handle the amplitude, we assume the density of rays is proportional to the flux through the surface. First, we generate a histogram over radial positions, then create an interpolating function from this, $\phi_\text{int}(r)$. In order to satisfy energy conservation and accommodate a spatially varying Poynting vector, we create an interpolation function for the ray angles, $\Theta_{int}(r)$. Finally, we estimate the electric field on a 2D grid $(x_i, y_j)$ as equation~\ref{eq LS field} . We choose grid spacing of $\qty{100}{\nano \meter} \sim \lambda / 4 n_\text{water}$, which is Nyquist sampled with respect to the maximum propagating spatial frequency.

\begin{equation}
    E(x_i, y_j) = \sqrt{\frac{\phi_\text{int}\left(\sqrt{x_i^2 + y_j^2} \right)}{\cos\left(\Theta_{int}\left(\sqrt{x_i^2 + y_j^2} \right)\right)}} \exp \left[i \ OPL_\text{int} \left(\sqrt{x_i^2 + y_j^2} \right) \right]
    \label{eq LS field}
\end{equation}

To determine the asymmetric light sheet intensity from these radially symmetric simulations, we combined two simulations using excitation beams having Gaussian waists matching the two asymmetric light sheet waists, $\omega_0^{(x)}$ and $\omega_0^{(y)}$, equation~\ref{eq asy gaussian}. We combine these two solutions in analogy to the following connection between intensities valid for Gaussian beams, equation~\ref{eq gaussian I}.
\begin{eqnarray}
    I_G(x,y,z;\omega_0^{i}) &=& \left(\frac{\omega_0^{i}}{\omega^{i}(z)}\right)^2 \exp \left[-2 \frac{\left(x^2 + y^2\right)}{\omega^{i}(z)^2}\right] \label{eq gaussian I}\\
    I_{asy}(x,y,z; \omega_0^{(x)}, \omega_0^{(y)}) &=& \frac{\omega_0^{(x)}\omega_0^{(y)}}{\omega^{(x)}(z)\omega^{(y)}(z)} \exp \left[\frac{-2 x^2}{{\omega^{(x)}(z)}^2} + \frac{-2 y^2}{{\omega^{(y)}(z)}^2}\right]
    \label{eq asy gaussian}\\
    &=& \sqrt{I_G \left(\sqrt{2}x,0,z; \omega_0^{(x)} \right)} \sqrt{I_G \left(0, \sqrt{2}y, z; \omega_0^{(y)}\right)}.
\end{eqnarray}
We assume a similar relationship holds for our model profiles $I_M$. In practice, we simulate the light sheet propagation along the axis with the tight light sheet focus, and assume an exact Gaussian form along the tall axis of the light sheet. We expect that optical aberrations only weakly affect this axis, and we find simulations using this approach agree well with our experiments.

\begin{equation}
    I_\text{LS}(x,y,z;\omega_0^{(x)},\omega_0^{(y)}) \approx \frac{\omega_0^{(x)}}{\omega^{(x)}(z)} \exp \left[-\frac{2x^2}{\omega^{(x)}(z)^2} \right] \sqrt{I_M \left(0, \sqrt{2} y, z;  \omega_0^{(y)} \right)} \label{eq LS intensity}
\end{equation}

Before determining the light sheet intensity profile, we first characterize the focal plane using ray tracing. Determining the focal plane in the presence of aberrations is complicated by the spread in optical axis intersection between paraxial and marginal rays. For small aberrations, the best-fitting Gaussian reference sphere is centered at the midpoint between the two ray intersections~\cite{born912}. Beyond this, diffraction effects must be considered. In our optical model, the focal plane is defined as the midpoint between the paraxial and marginal ray intersections with the optical axis. 

To characterize wavefront aberrations, we estimated the effective objective pupil wavefront by re-sampling the rays with an objective whose parameters depended on the resulting focal plane and immersion media. The action of the objective eliminated the curvature attributed to focusing, resulting a flat pupil plane to characterize wavefront aberrations. We defined the wavefront $W(r)$ as the optical path length difference with respect to the the paraxial ray. Wavefront distributions were fit using an $8^{th}$ order polynomial. The polynomial coefficients, $C_i$, are closely related to the Zernike aberration representation. Here we only include azimuthally symmetric aberrations, as our simulation is symmetric. For example, $C_4$ is a direct measure of spherical aberration while defocus is characterized by a combination of $C2$ and $C_4$\cite{wyant1992}. To quantify wavefront aberrations, we defined an aberration function by subtracting the piston term. For a holistic interpretation of the focus quality, we calculated the wavefront root-mean-square deviation (RMS) and Strehl ratio~\cite{born913, wyant1992}.

\begin{eqnarray}
    W(\rho) &=& \text{OPL}(\rho)-\text{OPL}(0) = \sum_{i=0}^{8} C_i \rho^i    
    \label{eq opl fit}\\
    \Phi(\rho) &=& W(\rho) - C_0
    \label{eq wf aber}\\
    S &=& \frac{1}{\pi^2} \left|\int_0^{2\pi} \int_0^1 e^{i k_o \Phi(\rho)} \rho  \ d\rho d\phi \right|^2
    \label{eq strehl}\\
    RMS &=& \sqrt{\frac{1}{\pi}\int_0^{2\pi}\int_0^1 \left| \Phi(\rho) - \overline{\Phi}\right|^2 \rho \ d\rho d\phi}
    \label{eq wf rms}
\end{eqnarray}
where $\rho$ and $\phi$ describe the normalized pupil position, and $\overline{\Phi}$ is the mean wavefront aberration.

Additionally, we employed ray tracing to characterize the axial spread of rays (fig.~\ref{fig remote_focus_comparison}). The difference in optical axis intersection between marginal and paraxial rays is a method to estimate spherical aberration~\cite{born53}.

\subsection*{Simulation of light sheet illuminations}

We simulated the remote focus pathway used in the light sheet acquisitions as shown in figure~\ref{fig model overview}A. Each model included a thick lens of variable focal length to simulate an electrotunable lens (ETL), two achromatic doublet lenses ($L_1$, $L_2$) and an objective ($EXC_{obj}$). We assume the light sheet propagates through a cuvette, which we model as a flat (plano-plano) glass surface of thickness \qty{1.25}{\milli\meter} and RI=$1.4585$. The region inside the cuvette is filled with water, RI=$1.33$. The RI mismatch at the cuvette surface shifts the focus and reduces the effective NA.

The ETL was modeled using a flat first surface, a spherical second surface of variable curvature and fixed edge thickness. The ETL was parameterized by the edge thickness and the lens power in diopters (\unit{\meter^{-1}}). The lens material RI was set to 1.3 to match the manufacturers specifications, and edge thickness was estimated to be \qty{10}{\milli\meter}. To map the ETL power or focal length to the second surface radius of curvature, $R_2$, we applied the lens makers equation for a plano-convex or concave lens.

To align the lenses and ensure the ETL and objective are in conjugate planes, we deployed ABCD matrix methods to identify the lens' principle and focal planes. After ray tracing each component, we determined the intensity distribution about the focus as described above and extracted the central slice, $I_M(x, 0, z)$ for comparison to experimental data.

All computation, except for data acquisition, was performed using a Linux Mint 20 server with two Xeon E5-2650 CPUs (Intel) with \num{16} cores each, \qty{1}{\tera \byte} RAM, and two GeForce RTX 3090 Ti (Nvidia) with \qty{24}{\giga \byte} of memory.

\subsection{ETL remote focus optimization}

To investigate the effects of using an ETL for remote focusing we used ray tracing to characterize the axial aberrations. The remote focus pathway includes the same components as described above, with and without the cuvette and with models for multiple commercial objectives (0.14 NA Mitutoyo 378-802-6, 0.3 NA Nikon CFI Plan Fluor MRH00105). In each case we tested all physically attainable configurations of cuvette position and ETL curvature. The cuvette offset positions spanned the space between the lenses surface element and the lenses working distance. In the presence of the cuvette, the ETL power is constrained to keep the focal plane within the cuvette bounds. The light sheet displacement or focal shift is calculated with respect to the last cuvette surface. 

\subsection*{Experimental Setup}
\begin{figure}[ht!]
    \centering
    \includegraphics[width=1.0\textwidth]{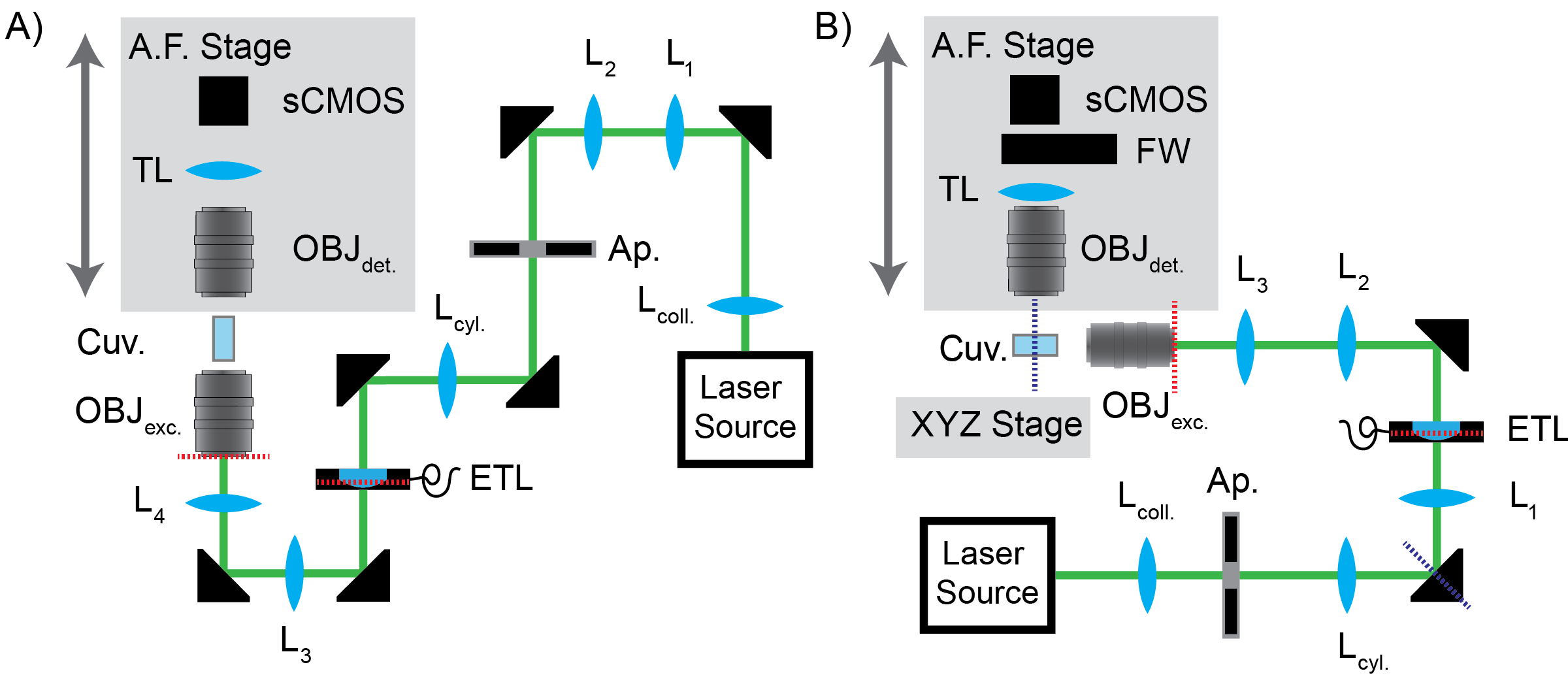}
    \caption{(A) Optical pathway used to image light sheets in transmission. (B) Optical pathway used to characterize PSFs to validate ETL optimization model. $L_i$ are achromat doublets, \textit{Ap.} is the adjustable aperture used to modulate the NA. $L_{cyl}$ is the cylindrical lens, \textit{TL} is the tube lens. The dotted line represents conjugate planes, red and blue indicating conjugate to the $\text{BFP}_{EO}$ and $\text{FFP}_{EO}$ respectively.}
    \label{fig imaging setup}
\end{figure}

We modified an existing remote focus SPIM setup to capture light sheet images in transmission~\cite{ryan2017,singh2017}, as illustrated in Figure~\ref{fig imaging setup}A. To validate the ETL optimization model, we imaged beads embedded in agar using the optical pathway in Figure~\ref{fig imaging setup}B. A National Instruments DAQ (USB-6343) was used to synchronize the ETL and illumination. Below, we detail the illumination pathway, followed by the detection optics.

Both excitation optical pathways utilized an excitation source (Oxxius L4Cc) with four wavelengths (\qty{488}{\nano \meter}, \qty{523}{\nano \meter}, \qty{561}{\nano \meter}, \qty{638}{\nano \meter}) combined into a single fiber output, with \qty{561}{\nano \meter} used for light sheet measurements and \qty{488}{\nano \meter} used to image beads. For all measurements, the fiber output was collimated using \qty{250}{\milli \meter} focal length achromatic lens (Thorlabs AC508-250-A-ML), the effective NA was modulated using an adjustable rectangular slit (Ealing 74-1137-000) conjugate to the $\text{BFP}_{EO}$, and an ETL (Optotune EL-16-40-TC) was utilized for the remote focusing unit. 

In the excitation optical pathway used for light sheet characterization, Fig.~\ref{fig imaging setup}A, the beam was collimated and expanded by a factor of 4 before passing through the aperture (Thorlabs AC508-250-A-ML, Thorlabs AC254-050-A-ML, AC508-200-A-ML) and a truncated Gaussian line profile was generated in the $\text{BFP}_{EO}$ by focusing a cylindrical lens (Thorlabs ACY254-200-A) on the ETL and relayed to the $\text{BFP}_{EO}$ (Thorlabs AC508-150-A-ML, AC508-100-A-ML) before being focused by a 0.3 NA air objective (Nikon MRH00105). 

The excitation optical pathway used for standard SPIM operation used a cylindrical lens (ACY254-50-A) to generate a line profile conjugate to the imaging plane. The line profile is refocused on the ETL and relayed to the $\text{BFP}$ of a 0.14 NA air objective (Thorlabs AC254-050-A-ML, AC508-180-A-ML, AC508-150-A-ML, Mitutoyo 378-802-6).

For both measurements, the detection optics were mounted on an adjustable motorized stage (Thorlabs KST101) to record z-stacks for light-sheet characterization or perform auto focusing during SPIM operation. Light sheets were imaged in transmission using a 0.5 NA long working distance objective (Mitutoyo 378-805-3) and a \qty{200}{\milli \meter} tube lens (Thorlabs TTL-200) focused on a scientific CMOS camera (Photometrics Iris15). The effective pixel size of \qty{85}{\nano \meter} was adequate to Nyquist sample all the light sheets considered here. For imaging beads, the detection optics were replaced with a 0.5 NA 2x magnification objective and 1x tube lens (Olympus MV PLAPO 2XC, Olympus MV PLAPO 1X), forming an image on the same scientific CMOS camera (Teledyne Photometrics Iris15). Fluorescence excitation was isolated using a \qty{30}{\milli \meter} filter wheel (Finger Lakes Instrumentation HS0433417) and barrier filter (Semrock FF01-536/40-32-D).

The control computer for all experiments was a Dell Precision 3630 running Microsoft Windows 10 Enterprise. This computer has a Core i7-9700K CPU (Intel) with \num{8} cores, \qty{64}{\giga \byte} RAM, and a GeForce RTX 3060 GPU (Nvidia) with \qty{12}{\giga \byte} of memory. The \textit{navigate} software package~\cite{marin2024} was used to collect the static and ASLM data shown in Fig.~\ref{fig lightsheet}A-B and Fig.~\ref{fig remote_focus_comparison}. Micromanager~\cite{edelstein2014} was used to acquire light sheets in transmission.

\subsection{Light sheet profile acquisitions}

We measured light sheets focusing in air and through an RI mismatch as expected during normal imaging conditions. The RI mismatch was created using a  \qtyproduct{10 x 20 x 45}{\milli \meter} (\qty{22.5}{\milli \meter} path length) glass cuvette filled with water or $1\%$ agar. For each, we modulated the effective NA by adjusting the long axis extent of the laser elliptical focus in the objective back focal plane then acquired a z-stack for a range of ETL powers. The position of the index mismatch is fixed such that when the ETL is flat, the focus is centered in the cuvette. The remote focus distance is measured relative to the flat ETL's light sheet position. After adjusting the aperture, the ETL current was reset to zero, and the detection stage adjusted to center the light sheet. To acquire data, we first moved the detection stage to align the FOV to the desired remote focus distance, then adjusted the ETL power to recenter the light sheet focus. For each NA, we captured z-stacks for remote focus positions ranging from \qty{\pm 4.0}{\milli \meter}. The remote focus range for the higher effective NAs was limited by aberrations. The z-stack range was adjusted to observe out of focus features of the light sheets.

The 3D raw data was processed to extract a light sheet ``slice" which we could compare to our simulation results. First, we identified the focal plane using the peak intensity. The light sheet slice or ZX plane with the greatest average intensity was chosen to compare with our model results. The resulting 2D slice image is rotated to align the propagation and $z$-axis, typically less than \qty{1}{\degree}. For light sheets imaged in water, stage displacements were scaled by $1.33$ to reflect the image plane shift imposed by the index mismatch. 

\subsection{Light sheet profile analysis}

We quantified both the simulated and experimental light sheet's radial confinement or width, $\omega(z)$, and propagation length, $z_L$. The width is characterized by the full-width at half-maximum (FWHM) of the radial intensity and the length is the range over which $\omega(z) \leq \sqrt{2\ln{2}}\omega_0$.

To process raw images into light sheet parameters, We first interpolated the axial intensity to identify the diffraction focal plane. Next, the focal plane intensity was fit using a 3-Gaussian intensity model (eq.~\ref{eq:3gaussianfit}) to account for side lobes and more complex beam structure caused by diffraction. The width, $\omega_0$ was defined using the peak lobe width and reported using the FWHM. Next, we numerically evaluated the FWHM, including side lobes, for each plane. After calculating the width for each z-plane, we interpolated the resulting distribution and defined the length as the range over which the width increased by a factor of $\sqrt{2}$. In calculating the the light sheet length, we found a numerical approach to define the radial extents (where $\omega(z) \geq \sqrt{2}\omega_0$) was more consistent for comparing Gaussian and non-Gaussian light sheets than fitting using a Gaussian function. 

\begin{equation}
    \label{eq:3gaussianfit}
    I(r) = I_\text{offset} + \sum_{i=0}^{i=2} I_i \exp{\left[-2 \frac{(r-\mu_i)^2}{\omega_i^2}\right]} 
\end{equation}

\subsection*{Bead imaging to characterize light sheet quality}
We prepared \qty{1}{\mu \meter} beads embedded in \qty{1}{\percent} agar (FluoSpheres F8823). Using the standard SPIM setup, we performed imaging experiments at cuvette offsets positions from \qty{10}{\milli\meter} to \qty{35}{\milli\meter}, every \qty{5}{\milli\meter}. For each cuvette position, the detection stage was translated to focus \qty{\sim 1}{mm} from the leading surface, in the middle of the cuvette and \qty{\sim 1}{\milli\meter} from the back surface. We captured \qty{80}{\mu \meter} z-stacks at each position. To quantify our results, we localized and fit the beads to a Gaussian model~\cite{localize_psf} and then applied a median and mean filter to the resulting axial standard deviation distribution along the propagation axis. To estimate the propagation length of the light sheet, we numerically identified the FWHM of the dip about the focus.

\section*{Results}

\subsection*{Comparison of simulated and experimental light sheets}

\begin{figure}[ht!]
    \centering
    \includegraphics[width=1.0\textwidth]{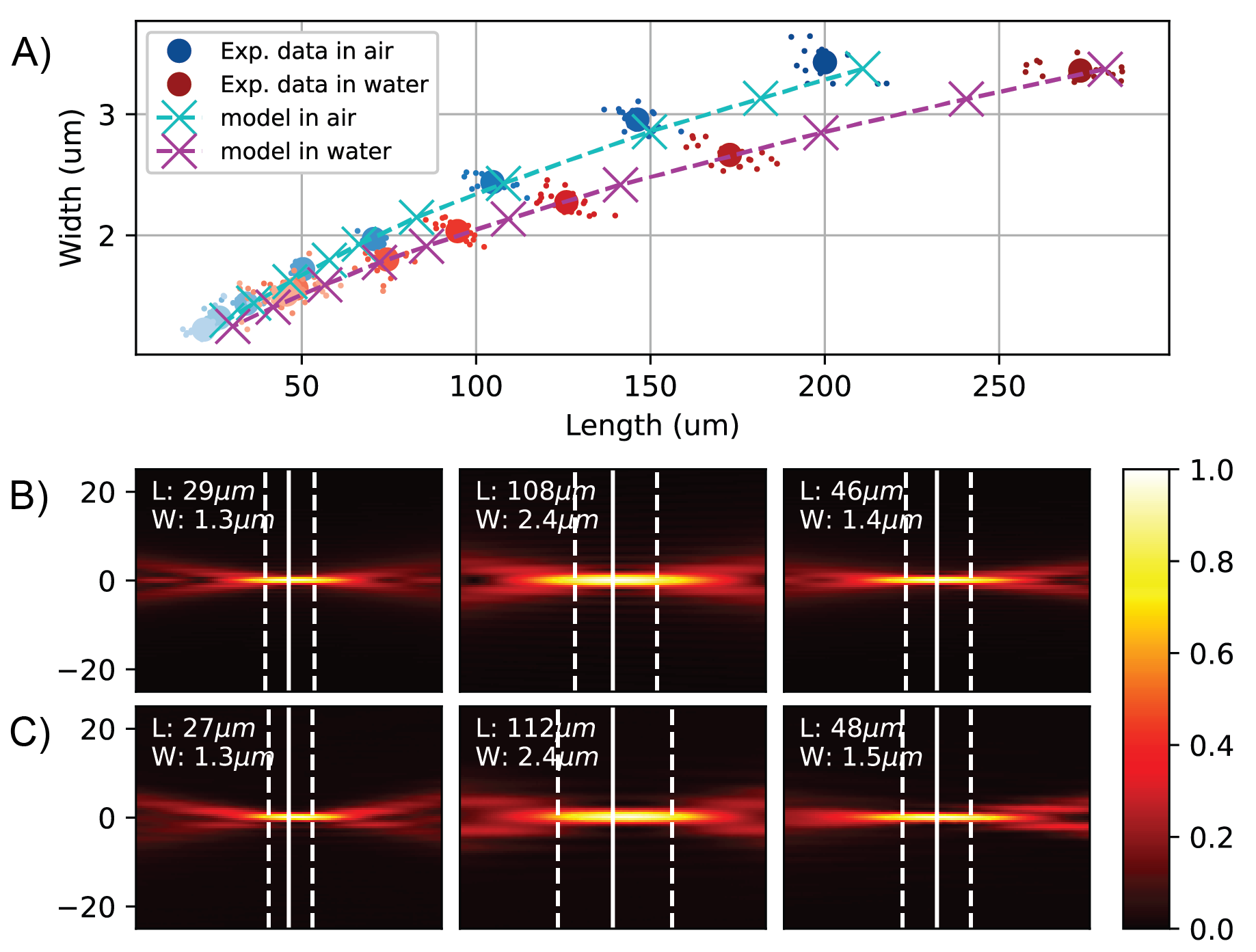}
    \caption{(A) Light sheet widths and lengths for simulated and experimentally light sheets. Visual comparison of various simulated (B) and experimental (C) light sheets. The first two columns are moderate and low NA examples focusing in air with no offset. The last column is an example of displaced light sheets. The focal plane (dashed) and light sheet edges (solid) are highlighted in white.
    }
    \label{fig lengths_vs_widths}
\end{figure}

The model results (dashed lines) depicted in Figure~\ref{fig lengths_vs_widths} represent light sheets with zero shift applied  are in agreement with prior results~\cite{chang2020,remacha2020}. The model results in water suggest an index mismatch is a limiting factor in obtaining tighter light sheets, but they also exhibit longer axial profiles. A comparison between the model results in air and in water supports prior theoretical work, which found that the spherical aberration induced by an index mismatch extends the propagation length compared to a Gaussian focus of equal width \cite{remacha2020,deng2021}.

The experimental results with no shift applied closely follow the model expectations from eq.~\ref{eq LS intensity} (Figure~\ref{fig lengths_vs_widths}). Our experimental results are consistent with previous observations of Gaussian beam or ``sinc" light sheets that exhibit longer propagation lengths for a given $\omega_0$~\cite{chang2020, liu2023}. Focusing in water, the tightest attainable light sheet is limited by aberrations. Both the model and experimental results show that focusing through an index mismatch is a limiting factor in achieving quality light sheets. In this initial experiment, we did not fine tuning the relative spacing of the excitation optics and cuvette for the RI mismatch case.

The visual comparison between the model and experimental data emphasizes our ability to accurately capture features of the light sheets (Fig.~\ref{fig lengths_vs_widths}B,C). While the model replicates focal features such as side lobes effectively, as aberrations increase, we observe small deviations for out-of-focus features. We suspected this discrepancy arises because these features are highly sensitive to the precise phase profile of the beam or to the objective lens aberrations, which we are not modeling. A visual comparison between experimental and acquisition light sheets with applied focus shows that we can capture the aberration characteristics associated with correcting first order defocus with an ETL. 

\subsection*{Remote Focus Optimization for Heterogeneous Refractive Index}

Using an objective lens designed for a different RI will lead to aberrations and impact both the excitation profile and detected image quality. In addition, many large sample SPIM experiments have a RI mismatch between the optics, sample immersion media, and sample itself leading to a displacement of the focus. Traditionally, the approach to addressing this issue involves holding the sample chamber constant and displacing the light sheet (either physically or using remote focusing) until it is within the sample. Leveraging our model, we explored optimal configurations for remote focusing in such environments.

\begin{figure}[ht!]
    \centering
    \includegraphics[width=1.0\textwidth]{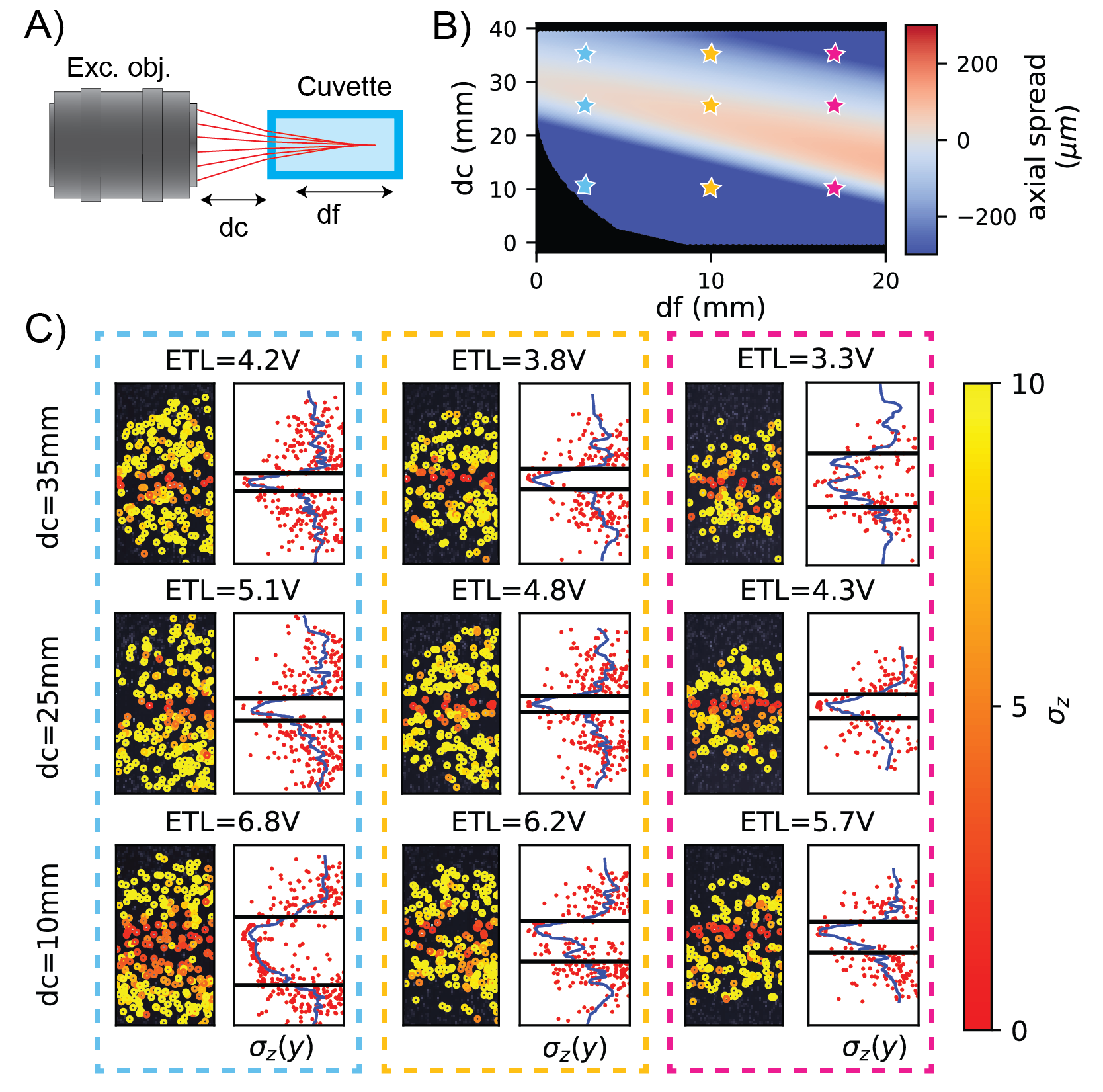}
    \caption{(A) The cuvette offset,  $dc$, is the distance between objective and RI mismatch. The light sheet displacement, $df$, is calculated with respect to the cuvette surface. (B) Axial aberration results for all possible combinations of $dc$ and $df$. Stars represent approximate position of experimental data. (D) Experimental static light sheet PSF results with z standard deviation, $\sigma_z(y)$ and focus extent (black lines). Each row represents a different cuvette position and each column corresponds to the same light sheet displacement.}
    \label{fig remote_focus_comparison}
\end{figure}

We find for a given imaging configuration, there are up to two optimal light sheet positions where the light sheet aberrations are minimized. This occurs because the ETL induces an optical path length difference ($\Delta \text{OPL}$) that counterbalances the path length difference accumulated through the remaining optics. The RI mismatch imposes an additional $\Delta \text{OPL}$, degrading focus quality and shifting the focal plane further from the objective ($\Delta \text{OPL} \propto \sqrt{\left(\frac{r}{d}\right)^2 + 1}-1$, where $d$ is the distance propagated in the medium and $r$ is the distance from the optical axis). Consequently, adjustments in the ETL lens curvature are required to maintain the focal position within the FOV. Using the ETL remote focus model, we predicted the optimal spacing between the objective and the cuvette to minimize the beam waist ($\omega_0$) and maintain these values when shifting the light sheet via remote focusing (Figure~\ref{fig remote_focus_comparison}).

We evaluate our predictions by imaging and quantifying the point spread function (PSF) across the camera FOV, using diffraction limited fluorescent micro-spheres in agar, for various light sheet shifts and objective to cuvette spacing. The PSF $\sigma_z$ fit results along the propagation direction reflect the optical sectioning attributed to the light sheet's focal region. Quantifying the extent over which the PSF remains confined in the light sheet propagation direction, we find that our model well predicts the optical sectioning and propagation length in the presence of RI mismatch induced aberrations. The optimal position for placing the cuvette to achieve a uniform PSF $\sigma_z$ within the propagation length and maximizing the remote focus range was determined to be between \qty{25}{\milli\meter} and \qty{30}{\milli\meter}, or approximately 70\% of the objective focal length. Furthermore, we observed that when the cuvette is positioned at either extreme—either too close or too far from the objective—the quality of the light sheet improves with displacements away from these extreme positions.

\section*{Discussion}

We have studied the effect of remote focusing and refractive index mismatch on light sheet intensity distributions for large FOV SPIM. Through numerical modeling and experimental validation, we confirmed that RI mismatches lead to aberrated light sheet intensity profiles. However, some recovery is possible by jointly optimizing the applied light sheet shift and the objective to sample chamber spacing.

Ray tracing to account for the remote focus pathway aberrations enabled us to estimate the remote focus range and determine the optimal RI mismatch position to maximize optical sectioning. This approach ensures that the light sheet focus is both optimized to enhance image contrast and correctly positioned within the sample. By incorporating remote focus pathway aberrations in the light sheet simulation, we demonstrated that considering both diffraction effects and non-paraxial aberrations is crucial for accurately characterizing light sheets in remote focus and RI mismatch applications.

Adaptive frameworks based on heuristic metrics or guide stars, which account for RI mismatches and other image-degrading aberrations, significantly improve obtainable image quality across multiple SPIM optical configurations~\cite{bourgenot2012, tomer2014, royer2016,ryan2017, hubert2019, chen2020, rai2022, li2022, rai2023, liu2023ad}. Our results suggest that simply starting with an optimized imaging configuration will enhance image contrast across the FOV. This finding extends to extended FOV methods, such as ASLM, where optimizing both the RI mismatch position and the light sheet position will further improve image quality, independent of additional corrective measures. Future work could extend these methods to account for invariant illumination in reconstructing light sheet acquisitions of large samples, where the cuvette position and remote focus parameters should be updated to maintain the best quality light sheet. Additionally, extending the model to the full image formation framework would enable exploration of adaptive optics informed by modeling to dynamically assist the correction of the illumination pathway aberrations.

Our findings on the optimal configurations for remote focusing, especially in the presence of RI mismatches, provide valuable guidelines for improving imaging quality in SPIM. We hope this study offers a practical strategy for enhancing SPIM performance across a range of biological samples and imaging configurations. 

%%%%%%%%%%%%%%%%%%%%%%%%%%  END body  %%%%%%%%%%%%%%%%%%%%%%%%%%

\section*{Funding}
SJS, PTB, and DPS acknowledge funding from NIH NHLBI (R01HL068702).

\section*{Disclosures}
The authors declare no conflicts of interest.

\section*{Data availability}
Example datasets for the imaging the light sheet in transmission and bead measurements are available~\cite{data2024}. All imaging data is available upon request. Code to run analysis is available on our Github repository~\cite{SPIMmodel2024}.

%%%%%%%%%%%%%%%%%%%%%%% References %%%%%%%%%%%%%%%%%%%%%%%%%

%%%%%%%%%% If using BibTeX:
% \bibliography{sample}
%\bibliographystyle{plain}
% \bibliography{references.bib}
%apsrev4-2.bst 2019-01-14 (MD) hand-edited version of apsrev4-1.bst
%Control: key (0)
%Control: author (8) initials jnrlst
%Control: editor formatted (1) identically to author
%Control: production of article title (0) allowed
%Control: page (0) single
%Control: year (1) truncated
%Control: production of eprint (0) enabled
%

\end{document}